\documentclass[twocolumn,showpacs,preprintnumbers,amsmath,amssymb]{revtex4}

\usepackage{graphicx}
\usepackage{dcolumn}
\usepackage{bm}

\begin{document}


\title{Generation of Fourier transform limited heralded single photons}

\author{Alfred B. U'Ren$^1$, Yasser Jer\'{o}nimo-Moreno$^1$, Jos\'{e} H. Garc\'{i}a-Gracia$^2$}
\affiliation{$^1$Divisi\'{o}n de F\'{i}sica Aplicada, Centro de
Investigaci\'{o}n Cient\'{i}fica y Educaci\'{o}n Superior de
Ensenada (CICESE), Baja California, 22860,
Mexico\\
$^2$Departamento de F\'{i}sica, Tecnol\'{o}gico de Monterrey, Nuevo
Le\'{o}n, 64849, Mexico}

\date{\today}

%
\newcommand{\epsfg}[2]{\centerline{\scalebox{#2}{\epsfbox{#1}}}}

\begin{abstract}
In this paper we study the spectral (temporal) properties of
heralded single photon wavepackets, triggered by the detection of an
idler photon in the process of parametric downconversion. The
generated single photons are studied within the framework of the
chronocyclic Wigner function, from which the single photon spectral
width and temporal duration can be computed.  We derive specific
conditions on the two-photon joint spectral amplitude which result
in both pure and Fourier-transform limited heralded single photons.
Likewise, we present specific source geometries which lead to the
fulfilment of these conditions and show that one of these geometries
leads, for a given pump bandwidth, to the temporally shortest
possible heralded single photon wavepackets.
\end{abstract}

\maketitle


\section{Introduction}

Single photon wavepackets are a crucial ingredient for a number of
quantum-enhanced technologies such as quantum computation with
linear optics\cite{kok05} and secure quantum key
distribution\cite{gisin02}. Recently it has been shown that single
photons can be efficiently ``heralded'', whereby photon pairs are
generated by the process of spontaneous parametric downconversion
(PDC) and subsequently one of the photons is detected, thus
conditionally preparing a single photon wavepacket in the conjugate
mode\cite{uren04,uren05b,pittman05,alibart04,fasel04}.  While great
progress has been made in the development of so-called on demand
single photon sources (e.g. based on quantum dots in
micro-cavities\cite{michler00}, single atoms coupled to optical
cavities\cite{kuhn02} and nitrogen vacancies in
diamond\cite{kurtsiefer00}), PDC can lead to practical,
room-temperature sources of heralded single photons. A crucial
property of PDC-based heralded single photon sources is that the
spatial and spectral emission properties can be controlled to a
large degree by appropriate choices of the source geometry,
including dispersion characteristics of the nonlinear crystal, and
the pump characteristics.   It is indeed remarkable that PDC can be
configured to produce photon pairs ranging from those which are
factorable to those which are highly entangled in continuous degrees
of freedom such as frequency and transverse wavevector.  In this
paper we exploit this flexibility to derive conditions for the
generation of both pure \textit{and} Fourier transform limited
heralded single photons. Likewise, we present specific source
geometries which lead to the fulfilment of such conditions.  Thus,
the techniques presented here can be used to design PDC photon pair
sources leading to heralded single photons described by FT-limited,
ultrashort wavepackets.

Heralded single photon sources rely on the quality of the
photon-number correlations between the signal and idler (used as
trigger) PDC modes. It is experimentally straightforward to
determine such quality by measuring the ratio of the coincidence
detection rate across the signal and trigger PDC modes to the signal
singles count rate. There are a number of experimental factors which
can reduce this ratio from its ideal unity value.   Higher-order PDC
contributions whereby more than one photon pair can be emitted at a
given time implies that a trigger detection event (from a non-photon
number resolving detector) can erroneously indicate the presence of
a single signal photon, while in reality two, or more, photons are
present. Likewise, the presence of optical losses in the signal mode
implies that a trigger detection event can erroneously indicate the
presence of a single signal photon, while in reality a vacuum state
exists. The presence of background photons, for which a trigger is
detected without the existence of a signal photon evidently also
reduces the heralding efficiency.   In Ref.~\cite{uren05b} we have
proposed and implemented an experimental criterion which assesses
source performance, specifically suited to heralded single photon
sources, including the effects of optical losses, higher photon
number contributions, background photons and the binary behavior of
realistic single photon detectors.\\
\indent Apart from the quality of photon number correlations between the two
PDC modes, it is important to consider how entanglement in
continuous degrees of freedom (e.g. spectral and transverse
wave-vector) translates into the properties of the heralded single
photon. Specifically, as was shown in Ref.~\cite{uren05a},
correlations in any degree of freedom in the photon pairs results in
a departure from purity in the heralded single photons.   Such
impurity is undesirable as it implies that multiple single photons
produced by more than one source will not interfere. In this paper
we further analyze the spectral (temporal) properties of heralded
single photon wavepackets, showing for a restricted class of states
that factorable photon pairs lead to both \textit{pure} and
\textit{Fourier transform limited} heralded single photons. We show
in addition that in a specific source geometry the heralded single
photons can attain their shortest possible temporal duration, which
corresponds to the pump pulse duration.\section{Description of heralded single photons}

The heralded single photon in the signal mode can be represented by
its density operator $\hat{\rho}_s$:

\begin{equation}
\hat{\rho}_s=\mbox{Tr}_i(\hat{\rho} \hat{\Pi}_t),\label{E:2photrho}
\end{equation}
where $\mbox{Tr}_i$ represents a partial trace over the trigger
mode, in terms of the density operator of the two photon state
$\hat{\rho}$ and of the measurement operator $\hat{\Pi}_t$ for the
trigger (idler) mode:

\begin{equation}
\hat{\Pi}_t=\int d \omega g(\omega) |\omega\rangle_{i} \langle
\omega |_i,
\end{equation}
which has been expressed in terms of the frequency-dependent trigger
detection efficiency $g(\omega)$.

\begin{figure}[h]
\includegraphics[width=3in]{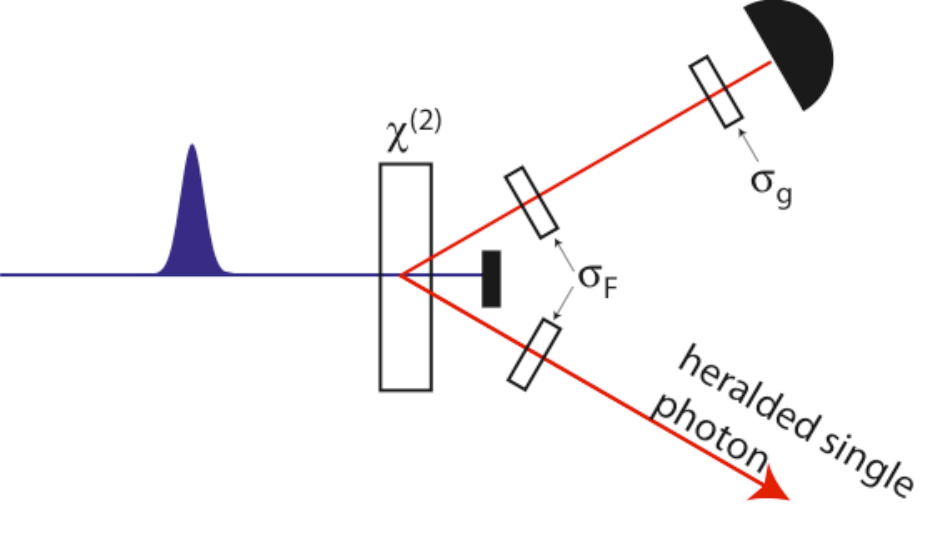} \caption{(color online) Heralded single photon
source, based on pulsed parametric downconversion in a $\chi^{(2)}$
crystal.\label{Fig:setup}}
\end{figure}

We will consider a heralded single photon source based on the
process of parametric downconversion (PDC) pumped by an ultrashort
pulse, shown schematically in Fig.~\ref{Fig:setup}.  The two-photon
state for the PDC process assuming specific directions of
propagation (i.e. disregarding the transverse wavevector degree of
freedom) can be expressed as:

\begin{equation}
|\Psi\rangle=\int\int d \omega_s d\omega_i f(\omega_s,\omega_i)
|\omega_s\rangle_s |\omega_i\rangle_i,\label{E:2photstate}
\end{equation}
where
$|\omega_\mu\rangle=\hat{a}^\dag(\omega_\mu)|\mbox{vac}\rangle_\mu$
with $\mu=s,i$ in terms of the joint spectral amplitude (JSA):

\begin{equation}
f(\omega_s,\omega_i)=\Phi(\omega_s,\omega_i)
\alpha(\omega_s+\omega_i) F(\omega_s,\omega_i),\label{E:JSA}
\end{equation}
which is in turn given by the product of the phasematching function
(PMF) $\Phi(\omega_s,\omega_i)$, the pump envelope function (PEF)
$\alpha(\omega_s+\omega_i)$ and a filter function
$F(\omega_s,\omega_i)$. While the PMF contains information about the
optical properties of the nonlinear crystal, the PEF characterizes
the pump field utilized. The function $F(\omega_s,\omega_i)$
describes the action of identical spectral filters (labelled
$\sigma_F$ in the figure) acting on both generated photons . The JSA
is assumed to be normalized so that the integral over both frequency
arguments of its absolute value squared yields unity. Substituting
the two-photon density operator $\hat{\rho}=|\Psi\rangle \langle
\Psi |$ into Eq.~\ref{E:2photrho}, we arrive at the following
expression for the density operator which results for the heralded
single photon in the signal mode:

\begin{equation}
\hat{\rho}_s=\int\limits_{0}^\infty d \omega_0 g(\omega_0)
 |\phi(\omega_0)\rangle \langle \phi(\omega_0)|.
\label{E:rho}\end{equation}
where
\begin{equation}
|\phi(\omega_0)\rangle=\int\limits_{0}^\infty
d\omega_1f(\omega_1,\omega_0)|\omega_1\rangle_s.\label{E:phi}
\end{equation}

Let us note than, in contrast to the two-photon density operator
$\hat{\rho}=|\Psi\rangle \langle \Psi |$, in general the single
photon density operator $\hat{\rho}_s$ represents a mixed state. In
order to gain physical insight, let us consider the case where
spectral filtering is applied to the trigger mode [such a filter is
labelled $\sigma_g$ in Fig.~\ref{Fig:setup}]. In the limit of strong
filtering we may write the trigger detection efficiency as a delta
function $g(\omega)\rightarrow \delta(\omega-\omega_F)$. Under these
circumstances, the resulting single photon density operator is given
by:
\begin{equation}
\hat{\rho}_s=|\phi(\omega_F)\rangle \langle\phi(\omega_F) |,
\end{equation}
which evidently represents a pure state\cite{rubin00}. Thus, an
avenue towards a pure heralded single photon source is to implement
strong spectral filtering on the trigger mode.  However, this is
achieved at the cost of a reduction in source brightness as well as
a reduction in the bandwidth of the prepared photon through
non-local state projection.

Is it possible to obtain a pure state without resorting to strong
spectral filtering?  Let us consider the case where the joint
spectral amplitude $f(\omega_s,\omega_i)$ is factorable i.e. where
functions $p(\omega)$ and $q(\omega)$ exist such that
$f(\omega_s,\omega_i)=p(\omega_s) q(\omega_i)$.  It is
straightforward to show from Eqns.~\ref{E:rho} and \ref{E:phi} that
under these circumstances the heralded single photon is indeed pure.
In order to gain understanding of the conditions for single photon
purity, let us perform a Schmidt decomposition of the two photon
state. The JSA can now be expressed as:
\begin{equation}
f(\omega_s,\omega_i)=\sum_m \sqrt{\lambda_m} u_m(\omega_s)
v_m(\omega_i),\label{E:Schmidt}
\end{equation}
where $u_m(\omega)$ and $v_m(\omega)$ represent the Schmidt
functions and $\lambda_m$ represents the Schmidt
eigenvalues\cite{cklaw00}. The resulting two photon density operator
$\hat{\rho}_s$ can be expressed as:
\begin{equation}
\hat{\rho}_s= \sum_m\sum_n \sqrt{\lambda_m \lambda_n}\int d\omega_0
g(\omega_0) v_m(\omega_0)v_n^*(\omega_0) |\phi_m\rangle \langle
\phi_n|, \label{E:rhoSchmidt}
\end{equation}
in terms of the Schmidt single photon wavepackets:
\begin{equation}
|\phi_m\rangle=\int d\omega u_m(\omega) |\omega
\rangle_s.\label{E:SchmidtWavePacket}
\end{equation}

It is evident from Eq.~\ref{E:rhoSchmidt} that in general the
heralded single photon is described by a mixed state, given as an
incoherent sum of the Schmidt wavepackets.  This expression
simplifies considerably in the case where the trigger detection
efficiency is not frequency dependent; using orthogonality of the
Schmidt functions, Eq.~\ref{E:rhoSchmidt} then reduces to:
\begin{equation}
\hat{\rho}_s= \sum_m \lambda_m  |\phi_m\rangle \langle \phi_m|.
\label{E:rhoSchmidtSimp}
\end{equation}

From an analysis of Eqns.~\ref{E:rhoSchmidt} and
\ref{E:rhoSchmidtSimp} it becomes clear that if the two photon state
contains a single pair of Schmidt modes, i.e. if there is a single
term in the sum in Eq.~\ref{E:Schmidt}, the resulting heralded
single photon state is pure.  Thus, by engineering the state
\textit{at the source} so as to guarantee factorability (for which
the state is expressed in terms of a single Schmidt mode pair) it
becomes possible to achieve pure single photon conditional
generation without resorting to spectral filtering.  Conversely, for
a progressively larger number of active Schmidt mode pairs in the
photon pair, the attainable single photon purity decreases.  Indeed,
as was shown in Ref.~\cite{uren05a}, the single photon purity
$p=\mbox{Tr}(\hat{\rho}_s^2)$ is given by $p=1/K$ where the Schmidt
number $K$ is a measure of the number of active mode pairs.  In this
case, a pure heralded single photon may still be obtained by
spectral filtering at the cost of a reduction in both single photon
bandwidth \textit{and} signal level. Thus, factorable photon pair
generation represents a crucial enabling technology for future
progress in experimental quantum information processing.

\section{Chronocyclic Wigner function}

It is convenient to study the temporal and spectral properties of
heralded single photons within the framework of the chronocyclic
Wigner function (CWF) formalism.  The latter represents a useful
tool for the description of classical ultrashort
pulses\cite{paye92,walmsley96} and it has also been used to
represent the PDC two-photon state\cite{grice97th}. In this paper we
consider the CWF for the heralded single photon, which can be
expressed in terms of the single photon density operator
$\hat{\rho}_s$ as:
\begin{equation}
W_s(\omega,t)=\frac{1}{2 \pi}\int\limits_{-\infty}^\infty d \omega'
\langle \omega+\omega'/2|\hat{\rho}_s|\omega-\omega'/2\rangle
\mbox{e}^{i \omega' t}. \label{E:CWF}
\end{equation}
where subscripts $s$ in the frequency bra and ket have been omitted
for clarity. Such a function fully characterizes the spectral and
temporal properties of the single photons; indeed, the marginal
distribution resulting from integrating the CWF over frequency or
time results in the temporal or spectral, respectively, single
photon intensity profile. In terms of the joint spectral amplitude
$f(\omega_s,\omega_i)$, the chronocyclic Wigner function for the
heralded single photon can be expressed as:
\begin{eqnarray}
W_s(\omega,t)&=&\frac{1}{2 \pi}\int d\omega_0 g(\omega_0) \int d
\omega' f\left(\omega+\frac{\omega'}{2},\omega_0 \right)\times
\nonumber\\
&&f^*\left(\omega-\frac{\omega'}{2},\omega_0 \right) e^{i \omega'
t}.\label{E:CWF2}
\end{eqnarray}

Let us note that the chronocyclic Wigner function, as expressed in
Eq.~\ref{E:CWF2}, is  a generalization of the Wigner-Ville
function\cite{paye92} for a spectral amplitude function
$f(\omega_1,\omega_2)$ which depends on two rather than one
variable; the additional integration over the second frequency
argument results from tracing over one of the two photons in the
heralding process.

The spectral intensity profile of the heralded single photons is
given by the integral over the time variable of the chronocyclic
Wigner function:
\begin{equation}
I_\omega(\omega)=\int dt W_s(\omega,t) =\int d\omega_0 g(\omega_0)
|f(\omega,\omega_0)|^2,
\end{equation}
while the temporal intensity profile is given by the integral over
the frequency variable of the chronocyclic Wigner function:
\begin{eqnarray}
I_t(t)&=&\int d\omega W_s(\omega,t) \nonumber \\
 &=&\frac{1}{2 \pi}\int d \omega_0
g(\omega_0) \left|\int d \omega f(\omega,\omega_0) e^{i \omega
t}\right|^2.
\end{eqnarray}

In terms of the Schimdt functions, we may express the CWF as:
\begin{eqnarray}
W_s(\omega,t)&=&\frac{1}{2 \pi}\sum\limits_m\sum\limits_m
\sqrt{\lambda_m\lambda_n}\int d \omega_0 g(\omega_0) v_m(\omega_0)
v_n^*(\omega_0) \nonumber \\
&&\times\int d \omega'
u_m(\omega+\frac{\omega'}{2})u_n^*(\omega-\frac{\omega'}{2})e^{i
\omega' t}.\label{E:WignerSchmidt}
\end{eqnarray}

If the trigger detection efficiency is not frequency dependent,
using orthogonality of the Schmidt functions,
Eq.~\ref{E:WignerSchmidt} reduces to:
\begin{eqnarray}
W_s(\omega,t)=\frac{1}{2 \pi}\sum\limits_m \lambda_m\ \int d \omega'
u_m(\omega+\frac{\omega'}{2})u_m^*(\omega-\frac{\omega'}{2})e^{i
\omega' t}.\label{E:WignerSchmidtSimp}
\end{eqnarray}

In this case, the spectral and temporal intensity profiles are given
by:
\begin{eqnarray}
I_\omega(\omega)&=&\sum\limits_m \lambda_m
|u_m(\omega)|^2,  \\
I_t(t)&=&2\pi\sum\limits_m \lambda_m |\tilde{u}_m(t)|^2,
\end{eqnarray}
where $\tilde{u}_n(t)$ represents the Fourier transform of
$u_n(\omega)$.  In order to carry out more explicit calculations, we
will assume PDC photon pairs within a restricted, though quite
general, class of states.  We will consider a frequency-degenerate
collinear PDC source (centered at frequency $\omega_c$) based on
type-II phasematching, pumped by a short pulse (centered at
frequency $2 \omega_c$)  and will disregard group velocity
dispersion (as well as higher-order dispersion) terms. The
phasematching function for such an interaction can be expressed as
follows:
\begin{equation}
\phi(\omega_s,\omega_i)=\mbox{sinc}[L \Delta k(\omega_s,\omega_i)/2]
\mbox{exp}[i L \Delta k(\omega_s,\omega_i)/2],\label{E:PMF}
\end{equation}
with the phasemismatch expressed as a Taylor series up to first
order:
\begin{equation}
L \Delta k(\omega_s,\omega_i)\approx L \Delta k^{(0)}+\tau_s
(\omega_s-\omega_c)+\tau_i(\omega_i-\omega_c),
\end{equation}
in terms of the PDC central frequency $\omega_c$, the constant term
of the Taylor series $\Delta k^{(0)}$ (which we assume to vanish)
and the longitudinal temporal walkoff terms between the pump pulse
and each of the signal and trigger photons:
\begin{eqnarray}
\tau_s&=&L(k_p'(2 \omega_c)-k_s'(\omega_c)), \\
\tau_i&=&L(k_p'(2 \omega_c)-k_i'(\omega_c)).
\end{eqnarray}

Let us note that by retaining only constant and group velocity
terms, we ignore temporal broadening effects resulting from
quadratic and higher order dispersion terms.  While evidently a
complete analysis should include these terms, for the case analyzed
in detail here which corresponds to collinear, degenerate type-II
PDC where the bandwidth of the generated light is somewhat
restricted (as compared, for example, to collinear, degenerate
type-I PDC), the group velocity approximation is justified.  We
model the pump envelope function as a Gaussian function with width
$\sigma$:
\begin{equation}
\alpha(\omega_s+\omega_i)=\exp\left[-\frac{(\omega_s+\omega_i-2
\omega_c)^2}{\sigma^2}\right].
\end{equation}

In order to obtain an analytic expression for the CWF in closed
form, we approximate the sinc function appearing in the
phasematching function (see Eq.~\ref{E:PMF}) as a Gaussian function
[$\mbox{sinc}(x)\approx \mbox{exp}(-\gamma x^2)$ with $\gamma
\approx 0.193$].  As we will discuss below, comparisons of the CWF
calculated numerically without resorting to approximations and the
analytic CWF suggest that the Gaussian approximation represents a
valid approximation in that it yields a useful indication of the
single photon bandwidth and temporal duration.   Likewise, we assume
a Gaussian shape (with central frequency $\omega_{g0}$ and bandwidth
$\sigma_g$) for the trigger detection efficiency $g(\omega)$ [see
Eq.~\ref{E:rho}]:
\begin{equation}
g(\omega)=\exp\left[-\frac{(\omega-\omega_{g0})^2}{\sigma_g^2}\right],
\end{equation}
and a Gaussian shape for the two-photon filter function
$F(\omega_s,\omega_i)$ with width $\sigma_F$:

\begin{equation}
F(\omega_s,\omega_i)=\exp\left[-\frac{(\omega_s-\omega_c)^2+(\omega_i-\omega_c)^2}{\sigma_F^2}\right].
\end{equation}

Carrying out the integrals we obtain a CWF of the following form:

\begin{eqnarray}
W_s(\omega,t)&=&\frac{1}{\pi \Delta \omega \Delta t}
\mbox{exp}\left(-\frac{[\omega-\omega_c+\Omega]^2}{\Delta
\omega^2}\right) \nonumber \\
&\times&\mbox{exp}\left(-\frac{[t-T]^2}{\Delta
t^2}\right),\label{E:CWFanalytic}
\end{eqnarray}
in terms of the temporal and spectral widths $\Delta t$ and $\Delta
\omega$,
\begin{eqnarray}
\Delta t&=&T_{ss},  \label{E:Dt}\\
\Delta \omega &=&\frac{T_{ii}}{\sqrt{T_{ss}^2 T_{ii}^2-T_{si}^4}}
\label{E:Dw},
\end{eqnarray}
temporal and spectral shift terms $T$ and $\Omega$:
\begin{eqnarray}
T&=&\tau_s/2, \\
\Omega&=&\frac{1}{\sigma_g^2}\frac{T_{si}^2}{T_{ss}^2
T_{ii}^2-T_{si}^4}(\omega_{g0}-\omega_{c}),\label{E:Omega}
\end{eqnarray}
where we have defined the quantities $T_{ss}$, $T_{ii}$ and $T_{si}$
as follows:
\begin{eqnarray}
T_{ss}^2&=&2/\sigma_F^2+2/\sigma^2+\gamma \tau_s^2/2, \label{E:Tss}\\
T_{ii}^2&=&1/\sigma_g^2+2/\sigma_F^2+2/\sigma^2+\gamma \tau_i^2/2, \label{E:Tii} \\
T_{si}^2&=&2/\sigma^2+\gamma \tau_s \tau_i/2.\label{E:Tsi}
\end{eqnarray}

It is straightforward to prove that for a two photon state for which
its corresponding joint spectral amplitude can be expressed fully in
terms of Gaussian functions, the spectral width $\Delta \omega$ and
temporal duration $\Delta t$ must satisfy the inequality $\Delta
t\Delta \omega \geq1$; here $\Delta \omega$ and $\Delta t$ represent
half widths at $e^{-1}$. Note that this is a particular instance of
the uncertainty relation $\delta t \delta \omega \ge 1/2$ which is
fulfilled for a broad class of functions (where $\delta$ refers to
the standard deviation). However, the two photon state is given in
terms of a sinc function, which does not have a well defined
variance and therefore we cannot apply this result. Our approach is
rather to use the Gaussian approximation in cases where it yields
essentially the same chronocyclic Wigner function as without
recourse to such an approximation, and employ the uncertainty
relation associated with Gaussian functions. In order to study
possible regimes resulting in Fourier transform-limited conditional
single photon generation, we compute the time-bandwidth product
$\mbox{TB}=\Delta \omega \Delta t$:
\begin{equation}
\mbox{TB}=\Delta t \Delta \omega=\left(1-T_{si}^4/T_{ss}^2 T_{ii}^2
\right)^{-\frac{1}{2}}.\label{E:TB}
\end{equation}

The quantity TB assumes its minimum value ($\mbox{TB}=1$) for
Fourier transform limited single photons.  Such photons are
characterized by the important property that their temporal duration
$\Delta t$ is the shortest possible compatible with a given spectral
width $\Delta \omega$. For single photons which are not Fourier
transform limited, the time-bandwidth product assumes a value
greater than unity; indeed, the specific value of $\mbox{TB}$
provides a convenient measure of the departure from the Fourier
transform limit. Clearly, the greater the numerical value of
$\mbox{TB}$, the longer the time duration $\Delta t =
\mbox{TB}/\Delta \omega$ for a given fixed spectral width.

Besides the temporal duration of the heralded single photon, another
important characteristic time to consider is the correlation time:
the time of emission \textit{difference} between the signal and
idler photons. Such a time is important because it defines the
uncertainty in the expected detection time of the heralded photon
relative to the detection of the trigger photon.  In order to
calculate the correlation time $\tau_c$, we first compute the joint
temporal intensity $|\tilde{f}(t_s,t_i)|^2$ where
$\tilde{f}(t_s,t_i)$ denotes the two-dimensional Fourier transform
of the joint spectral amplitude $f(\omega_s,\omega_i)$.  The joint
temporal intensity represents the probability density function for
generating a photon pair at specific times $t_s$ and $t_i$.  As a
second step, we may express the joint temporal intensity in terms of
the variables $t_{+}=t_s+t_i$ and $t_{-}=t_s-t_i$.  From this, we
can calculate the marginal probability distribution $S_{-}(t_{-})$
which yields the probability density function for generating a
photon pair with a certain time of emission difference $t_{-}$. For
the approximations already introduced in this paper, we may express
this $t_{-}$ marginal distribution as $S_{-}(t_{-})=\pi^{-1/2}
\tau_c^{-1}\mbox{exp}(-t_{-}^2/\tau_c^2)$ in terms of the
correlation time $\tau_c$:
\begin{equation}
\tau_c=\frac{\sqrt{8+\gamma \sigma_F^2(\tau_i-\tau_s)^2}}{\sqrt{2}
\sigma_F},\label{E:tauclong}
\end{equation}
which in the absence of spectral filtering ($\sigma_F \rightarrow
\infty$) reduces to:
\begin{equation}
\tau_c= \sqrt{\gamma/2} |\tau_{-}|,\label{E:tauc}
\end{equation}
where:
\begin{equation}
\tau_{-}=\tau_{s}-\tau_{i}=L[k_s'(\omega_c)-k_i'(\omega_c)].\label{E:taum}
\end{equation}

It is instructive to compare Eq.~\ref{E:Dt} (together with
\ref{E:Tss}) for the heralded single photon time duration with
Eq.~\ref{E:tauc} (together with Eq.~\ref{E:taum}) for the
correlation time. It becomes apparent that while the former is
defined by a group velocity mismatch term $\tau_s$ between the pump
and the signal photon, the latter is defined by a group velocity
mismatch term $\tau_{-}$ between the trigger and the signal photons.
It should also be stressed that the correlation time \textit{does
not} depend on the pump bandwidth\cite{friberg85}.

\section{FT-limited single photons via group velocity matching}

In order to make the discussion more specific we will now consider
the concepts developed so far for particular experimental
situations.   Let us first consider the case of PDC generation in
the limit of vanishing pump bandwidth (i.e. $\sigma \rightarrow 0$).
It is clear from Eqns.~\ref{E:Dt} and \ref{E:Tss} that, when the
pump bandwidth is much smaller than the reciprocal group velocity
mismatch term (i.e. $\sigma \ll 1/\tau_s$), the single photon time
duration becomes dependent only on the pump bandwidth:  $\Delta
t=\sqrt{2}/\sigma$. As might be expected this time duration tends to
infinity in the idealized CW-pump case.  Likewise, the spectral
width in the $\sigma \rightarrow 0$ limit can be computed from
Eq.~\ref{E:Dw} and yields $\Delta \omega=\sqrt{2/\gamma}/|\tau_{-}|$
[which in fact corresponds to the unfiltered reciprocal correlation
time (see Eq.~\ref{E:tauc})]. Thus, the time bandwidth product in
the limit of a very short pump bandwidth $\mbox{TB}=2/(\sqrt{\gamma}
\sigma |\tau_{-}|)$ diverges. We conclude therefore that CW-pumped
PDC light is ill-suited for generation of Fourier transform limited
heralded single photons.

\begin{figure}[h]
\includegraphics[width=3in]{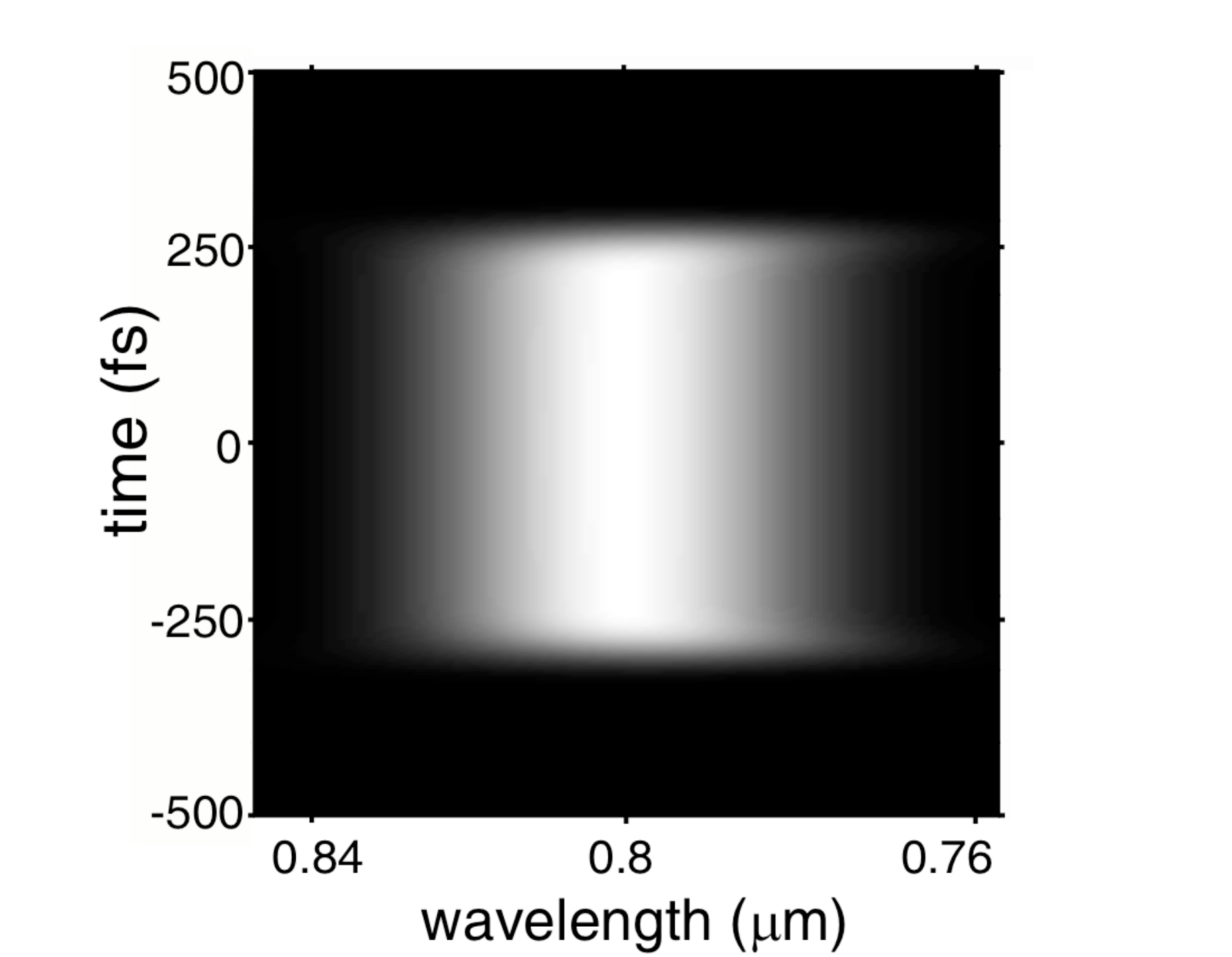} \caption{Chronocyclic Wigner function
(CWF) obtained numerically for a BBO crystal of $5$mm length,
phasematched for collinear, degenerate type-I PDC centered at
$800$nm, pumped by an ultrashort pulse train with a $5$nm bandwidth.
The resulting single photon time duration $\Delta t=288.01$fs and
single photon bandwidth $\Delta \omega=61.4$THz, with $\Delta t
\Delta \omega\approx17.7$, suggest a large departure from the
transform limit.\label{Fig:wignertype1}}
\end{figure}

Let us briefly consider degenerate, collinear type-I PDC as a
candidate for ultrashort heralded single photon generation.  Within
the first order treatment presented in this paper (ignoring group
velocity dispersion and higher order dispersive terms in the
phasemismatch $\Delta k(\omega_s,\omega_i)$), degenerate collinear
type-I photon pairs are characterized by $\tau_s=\tau_i$ or
equivalently by $\tau_{-}=0$. It has been shown\cite{uren05a} that
the orientation of the phasematching function in
$\{\omega_s,\omega_i\}$ space is given by $-\arctan(\tau_s/\tau_i)$,
so that in the case of degenerate collinear type-I PDC, this
orientation is fixed at $-45^\circ$. Furthermore, within the linear
approximation used, this function has an infinite extent along the
direction $\omega_s-\omega_i$ and shows perfect overlap with the
pump envelope function which is characterized by the same
orientation: i.e. all frequency pairs within the pump envelope
function are correctly phasematched.  Thus, within this
approximation the spectral width of each of the two single photon
wavepackets is infinite, while the temporal duration has a finite
value, determined by the pump bandwidth and crystal length. Let us
stress that the inclusion of GVD terms is necessary in the
degenerate, collinear type-I case in order to obtain a physically
meaningful result (while generally speaking the linear approximation
is justified for degenerate, collinear type-II PDC), leading to a
very large (but finite) spectral bandwidth for each photon.   The
very large spectral bandwidth (which becomes extraordinarily large
when the PDC photons are near the zero group velocity dispersion
wavelength\cite{zhang06}), however, unfortunantely does not result
in ultrashort single photon wavepackets (though it does imply a very
short correlation time, see discussion above and
Eqns.~\ref{E:tauclong} and \ref{E:tauc}).  In
Fig.~\ref{Fig:wignertype1} we present the chronocyclic Wigner
function (calculated numerically from Eq.~\ref{E:CWF2}) for a
specific collinear, degenerate type-I example. For a $5$mm long BBO
crystal (with a cut angle of $29.2^\circ$), pumped by an ultrashort
pulse train with $5$nm pump bandwidth producing photon pairs
centered at $800$nm we obtain $\Delta t=288.01$fs and $\Delta
\omega=61.4$THz, corresponding to $\Delta t \Delta \omega \approx
17.7$\cite{note} which suggests a large departure from the transform
limit. Apart from the above considerations, let us note that
degenerate, collinear type-I signal and idler photon pairs can at
best be split probabilistically leading to a maximum heralding
efficiency of $50\%$.  Thus, we likewise conclude that degenerate,
collinear type-I PDC is ill-suited for generation of Fourier
transform limited heralded single photons.

Let us turn our attention to the process of collinear, degenerate
type-II PDC pumped by ultrashort pulses.  In the propagation of
short optical pulses through dispersive media, group velocity
dispersion and higher dispersive terms result in temporal
broadening, while group velocity terms result in propagation without
distortion.  The situation is somewhat different for the generation
of PDC light by short pump pulses, where group velocity terms in the
nonlinear crystal can result in photon pairs where the constituent
single photon wavepackets are temporally much longer than the pump
pulse. We may write down the heralded single photon duration from
Eq.~\ref{E:Dt} in terms of the pump pulse duration
$\tau_p=\sqrt{2}/\sigma$ as:

\begin{equation}
\Delta t=\tau_p \sqrt{1+ \frac{2}{\sigma_F^2
\tau_p^2}+\frac{\gamma}{2}\frac{\tau_s^2}{\tau_p^2}}.\label{E:deltatexp}
\end{equation}

It is evident from Eq.~\ref{E:deltatexp} that the shortest possible
heralded single photon temporal duration is the pump pulse temporal
duration itself.  Likewise, we conclude from Eq.~\ref{E:deltatexp}
that this shortest temporal duration is obtained in the limit of no
spectral filtering ($\sigma_F \rightarrow \infty$) \textit{and} of
negligible group delay as compared to the pump pulse duration i.e.
$\tau_s \ll \tau_p$.  It becomes apparent that for an ultrashort
pump pulse, the attainable temporal duration is limited by the group
velocity mismatch term $\tau_s$. Two possibilities exist for making
$\tau_s /\tau_p \ll 1$.  The first involves using a very short
crystal, concretely one obeying $L\ll \tau_p/(k_p'-k_s')$; indeed
for a short enough crystal the group delay $\tau_s$ due to group
velocity mismatch can become insignificant. This involves the need
for progressively shorter crystals as the pump pulse duration is
decreased.  It clearly does not represent a viable solution as it
implies a prohibitive reduction in PDC flux. The second possibility
is to employ a PDC interaction obeying group velocity matching
between the pump and the PDC light. Indeed, it is clear from
Eq.~\ref{E:deltatexp} that for $\tau_s=0$ (which implies
$k_p'=k_s'$) we can obtain, with no spectral filtering, the shortest
possible time duration $\tau_p$ irrespective of the crystal length
used.

\begin{figure}[h]
\includegraphics[width=3.5in]{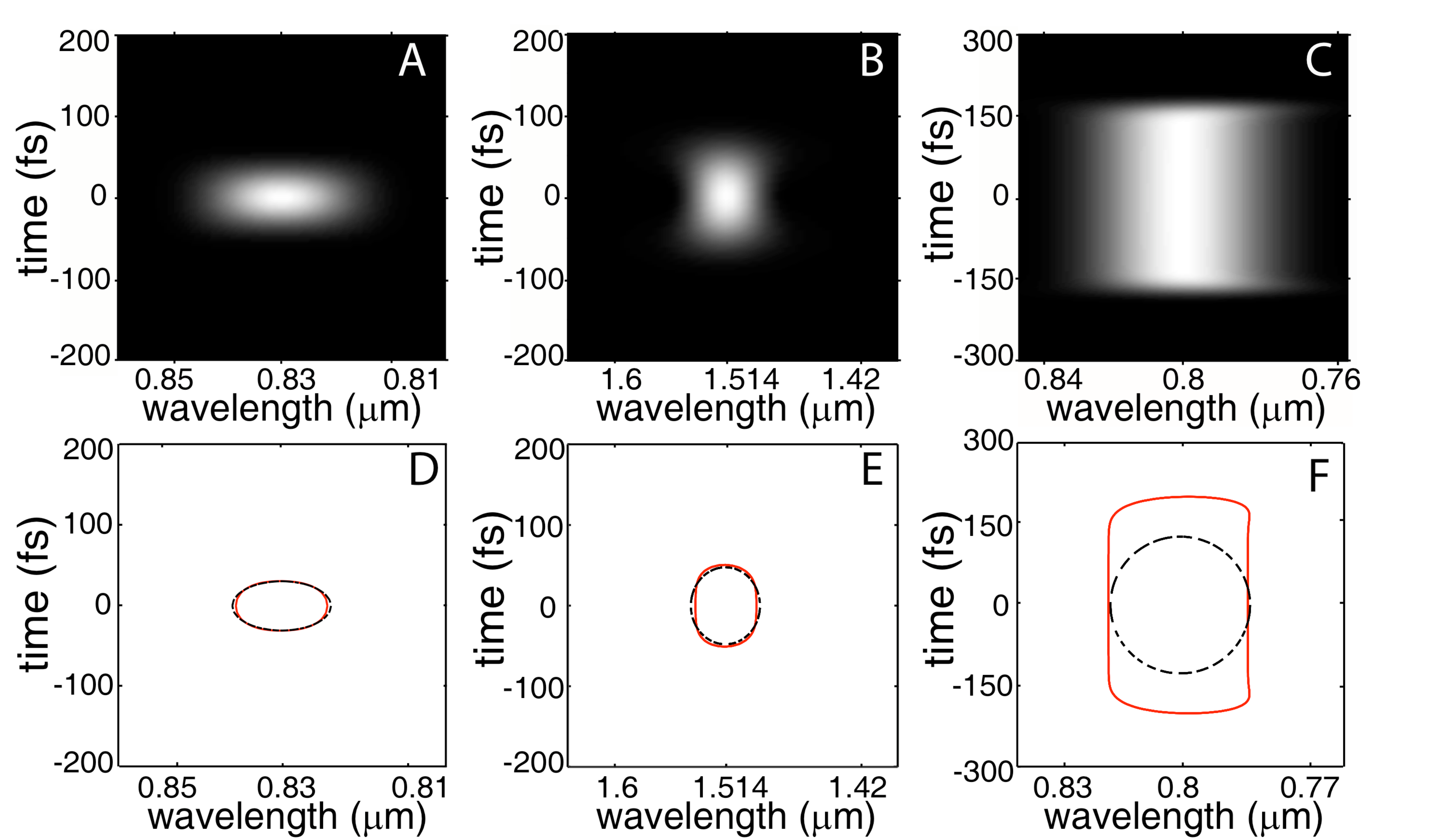}\caption{(color online) Panels
(A)-(C) show examples of numerically calculated chronocyclic Wigner
functions (CWF) without resorting to the Gaussian approximation and
taking into account all dispersive orders, centered at a time t=0,
for a single heralded single photon source based on: (A) a KDP
crystal, fulfilling asymmetric group velocity matching, (B) a BBO
crystal, fulfilling symmetric group velocity mathcing, and (C) a BBO
crystal not fulfilling any special group velocity condition.  Panels
(D)-(F) show, correspondingly to (A)-(C), the contour associated to
one half of the maximum value for the CWF calculated numerically
(solid line) and for the CWF calculated analytically (dashed line;
see Eq.~\ref{E:CWFanalytic}). \label{Fig:wigner}}
\end{figure}

Let us stress that in our analysis in this paper, we have ignored
group velocity dispersion terms (as well as higher order dispersive
terms).  For classical pulse propagation, under these circumstances
we would expect no temporal broadening.  However, Eq.~\ref{E:TB} for
the heralded single photon time-bandwidth product does indeed permit
values greater than unity, even for an unchirped pump. The physical
mechanism by which the single photon wavepackets which constitute a
PDC photon pair suffer temporal elongation is related to group
velocity mismatch between the pump pulse and the generated PDC
light. Indeed, in the long crystal limit the heralded single photon
duration is determined purely by the group velocity mismatch term:
$\Delta t_s \rightarrow \sqrt{\gamma} \tau_s / \sqrt{2}$.  In order
to understand this physically, let us consider a photon pair created
on the second face of the crystal and secondly a photon pair created
on the first face. While in the latter case the amplitudes for
photon pair generation emerge from the crystal temporally overlapped
with the pump pulse, in the former due to group velocity mismatch,
these amplitudes emerge ahead of the pump pulse by a time $\tau_s$.
Because the wavepackets corresponding to each of the two photons are
the result of adding the amplitudes from all locations within the
crystal, the result is a temporally stretched wavepacket (as
compared to the pump pulse). As a numerical example [to be discussed
further below; see Fig.~\ref{Fig:wigner}(C)], let us consider a
$5$mm long BBO crystal pumped by a pulse train centered at 400nm
with $5$nm bandwidth. The expected duration of the wavepackets
computed from Eq.~\ref{E:Dt} is $127.2$fs (for a pump pulse with
$28.3$fs duration), resulting in a time-bandwidth product of
$\mbox{TB}=6.4$. As has already been discussed, for small pump
bandwidths together with type-1 operation, the time bandwidth
product can take much larger values; indeed it can be shown that
values as high as $10^7$ for the Schmidt number $K$ and $\mbox{TB}$
are possible\cite{zhang06}.

How can we achieve Fourier transform limited conditional single
photon generation?  An analysis of Eq.~\ref{E:TB} reveals that
$\mbox{TB}=1$ if the correlation coefficient $T_{si}^2/(T_{ss}
T_{ii})$ vanishes.  One way in which this can occur is if strong
filtering is applied to both photons ($\sigma_F \rightarrow 0$) or
if strong filtering is applied to the trigger photon only ($\sigma_g
\rightarrow 0$).  In both of these cases, it can be readily shown
that in the limit of strong filtering, the Fourier transform limit
is attained (i.e. $\mbox{TB} \rightarrow 1$), evidently at the cost
of a prohibitive reduction in flux.  In addition, it should be
stressed that while the FT limit may be attained via filtering, the
bandwidth is thereby reduced thus precluding ultrashort
single-photon wavepackets. Alternatively, the correlation
coefficient $T_{si}^2/(T_{ss} T_{ii})$ vanishes if $T_{si}=0$ which
in fact represents the condition derived in Ref.~\cite{grice01} for
spectral factorizability in photon pair states.  Thus, we reach the
important conclusion that factorable two photon
states\cite{factorable,grice01,uren05a,kuzucu05} lead to Fourier
transform limited heralded single photons. Let us note that the
latter route involves no filtering, and therefore can lead to
bright, broadband sources of FT-limited single photons.

An analysis of the CWF [see Eq.~\ref{E:CWF}] reveals that the single
photon spectrum is in general shifted from the central PDC frequency
$\omega_c$ by a frequency $\Omega$ [given by Eq.~\ref{E:Omega}]
which depends linearly on the frequency detuning of the trigger
filter with respect to the central PDC wavelength.  Therefore, as
the trigger filter passband is shifted, the resulting heralded
single photon spectrum also shifts.  This is a direct consequence of
spectral correlation in the photon pair state; filtering of one of
the two photons, non-locally projects the conjugate photon into a
particular spectral band.  Indeed, $\Omega$ is proportional to
$T_{si}$, which controls the degree of factorizability in the photon
pair\cite{grice01}.  Thus, for a factorable two photon state (for
which $T_{si}=0$) the heralded single photon spectrum does not shift
due to shifts of the trigger filter passband. This suggests an
experimental test for the degree of factorizability (and thus for
the degree of departure from the FT limit): monitoring the heralded
single photon central frequency as a function of the trigger filter
central passband frequency.

The factorizability condition $T_{si}=0$ (see Eq.~\ref{E:Tsi}) can
be satisfied, for example, for the symmetric group velocity matching
(SGVM) condition\cite{keller97,grice01,kuzucu05}, expressed as
$\tau_s+\tau_i=0$, or alternatively as:

\begin{equation}
2 k_p'(2 \omega_c)-k_s'(\omega_c)-k_i'(\omega_c)=0.
\end{equation}

Physically, this condition implies that the signal and idler photon
amplitudes walk away longitudinally from the pump in a symmetric
fashion: i.e.  one outpaces the pump pulse by the same group delay
as its conjugate lags the pump pulse.  The resulting single photon
temporal duration can be computed from Eq.~\ref{E:Dt} together with
the conditions $\tau_s+\tau_i=0$ and $T_{si}=0$ (see
Eq.~\ref{E:Tsi}), obtaining the result

\begin{equation}
\Delta t=\sqrt{2} \tau_p.
\end{equation}

Thus, in the symmetric group velocity matching scenario, the single
photon time duration assumes, within a factor of $\sqrt{2}$, its
smallest possible value.  It may be similarly shown that the
spectral width is in this case simply the reciprocal of the temporal
duration so that the Fourier transform limit is attained without
recourse to filtering.

Let us study an alternative regime for Fourier transform limited
single photon generation, where the pump group velocity is matched
to the signal photon group velocity: i.e. $k_p'(2
\omega_c)=k_s'(\omega_c)$ or $\tau_s=0$, a regime which we refer to
as asymmetric group velocity matching (AGVM). In this case, it may
be readily shown that $\mbox{TB}\rightarrow 1$ if $\sigma \tau_i \gg
1$.  The latter condition may be re-expressed as:

\begin{equation}
\sigma L \gg \frac{1}{k_p'(2
\omega_c)-k_i'(\omega_c)}.\label{E:AGVMcond}
\end{equation}

Thus, by imposing the asymmetric group velocity matching condition
together with a large crystal length-pump bandwidth product, it
becomes possible to generate Fourier transform limited heralded
single photons. AGVM is preferable over SGVM in the sense that for
the former the heralded single photons attain the shortest possible
time duration $\Delta t=\tau_p$ (equal to the pump duration)
together with the largest FT-limited bandwidth possible $\Delta
\omega=1/\tau_p$ (while for the latter these parameters differ from
the ideal values by a factor of $\sqrt{2}$). However, it should be
stressed that AGVM leads to signal and idler photons which are
distinguishable from each other; the former is not a limitation for
heralded single photon generation but would be a limitation for
experiments where the signal and idler photons are made to interfere
with each other. Indeed, an advantage of the SGVM technique is that
it combines ultrashort FT-limited heralded single photon generation
with a symmetric joint spectral amplitude yielding indistinguishable
signal and idler photons (except in polarization).

Another important aspect that should be considered when designing a
heralded single photon source is the correlation time.  This time
determines the uncertainty in the expected time of arrival of the
heralded single photon with respect to the trigger detection time.
As has already been discussed, type-I PDC can result in extremely
short correlation times, due to the larger PDC bandwidths possible.
However, as has also been discussed, despite the short correlation
times, type-I leads to states exhibiting a strong deviation from the
FT limit.   Let us compare heralded single photon sources resulting
from SGVM and AGVM in terms of their correlation times. For SGVM,
$\tau_s=-\tau_i$ and therefore Eq.~\ref{E:tauc} reduces to
$\tau_c=\sqrt{2 \gamma}|\tau_i|$, while for AGVM $\tau_s=0$ and
therefore Eq.~\ref{E:tauc} reduces to
$\tau_c=\sqrt{\gamma/2}|\tau_i|$.  When comparing these two results,
it should be recalled that $\tau_i$ depends linearly on the crystal
length and that in order to obtain a factorable state through AGVM,
it is essential to use fairly long crystals [see
Eq.~\ref{E:AGVMcond}]. What this means in practice is that the
correlation time tends to be much longer for heralded single photons
produced through AGVM compared to heralded single photons produced
through SGVM.

Let us consider specific examples to illustrate the single heralded
photon duration and correlation time behavior.  First, let us
consider a $2$cm KDP crystal cut at $67.7^\circ$ (collinear,
degenerate type-II emission at $830$nm) which can yield FT-limited
single photons through AGVM\cite{uren05a}. Fig.~\ref{Fig:wigner}(A)
shows the CWF computed numerically for this case, assuming a pump
duration $\tau_p=30.4$fs (which corresponds to a $5$nm pump
bandwidth). The resulting single photon duration under the Gaussian
approximation is $\Delta t = \tau_p$ (where $\Delta \omega=1/\Delta
t$), while the correlation time is $896.3$fs. Secondly, let us
consider a $2.3$mm-length BBO crystal cut at $28.8^\circ$
(collinear, degenerate type-II emission at $1514$nm) which can yield
FT-limited single photons through SGVM\cite{grice01}.
Fig.~\ref{Fig:wigner}(B) shows the CWF computed numerically for this
case, assuming a pump duration $\tau_p=33.7$fs (which corresponds to
a $15$nm pump bandwidth). The resulting single photon duration under
the Gaussian approximation is $\Delta t = \sqrt{2} \tau_p$ (where
$\Delta \omega=1/\Delta t$), while the correlation time is $67.5$fs
(note this value is independent of the pump bandwidth used). Thus,
while AGVM leads to the shortest possible time durations, it
involves distinguishable signal and idler photons and longer
correlation times as compared to SGVM.

In Fig.~\ref{Fig:wigner}(C) we show, for comparison, the CWF
corresponding to a PDC source which does not fulfil any particular
group velocity condition and which therefore cannot be used as the
basis for an FT-limited heralded single photon source.  Here, we
have assumed a $5$mm-length BBO crystal cut at $42.3^\circ$ for
collinear, degenerate type-II emission at $800$nm. The resulting
single photon duration obtained numerically is $\Delta t=205.7$fs
while the single photon bandwidth is $\Delta \omega=50.1$THz,
yielding $\Delta t \Delta \omega\approx 10.3$\cite{note}, which
suggests a substantial deviation from the Fourier transform limit.
Figs.~\ref{Fig:wigner}(D)-(F) show the contour defined by
frequencies and times for which the CWF takes a value of $e^{-1}$
times the maximum value, calculated numerically (solid line) and
analytically through Eqn.~\ref{E:CWFanalytic} (dashed line) for each
of the examples in Figs.~\ref{Fig:wigner}(A)-(C). In general terms,
while the single photon bandwidth calculated numerically and
analytically agree very well, the analytic result based on the
Gaussian approximation tends to underestimate the temporal width. A
computation of the temporal widths calculated numerically $\Delta
t_{nu}$  and analytically $\Delta t_{an}$ while varying source
parameters, such as crystal length and pump bandwidth (not shown)
reveals that $\Delta t_{nu}/\Delta t_{an}$ ranges in value roughly
from $1$ to $1.6$. This is due to the difficulty of approximating
top-hat-type functions (which result from a sinc-type structure in
the spectral domain) with Gaussian functions. It should also be
pointed out, however, that there is a class of PDC two-photon states
for which the Gaussian approximation leads essentially to the same
result as predicted without resorting to this approximation.   This
includes those states for which the crystal length $L$ or the pump
bandwidth $\sigma$ (or both) are small.  Furthermore, the maximum
product $L\sigma$ for which the Gaussian approximation can be used
reliably depends on the orientation of the PMF in signal - idler
frequency space. Thus, for cases where the PMF is oriented
vertically such as the AGVM example above, the sinc structure is
fully contained by the idler single photon wavepacket, which is
traced over by heralding so that it has no effect on the prepared
signal photon wavepacket. The resulting agreement between the
numerical and analytic calculations  [see Fig.~\ref{Fig:wigner}(D)]
is essentially perfect; the very small disagreement is due to
quadratic and higher order dispersion terms which are ignored by the
analytic approach.  Likewise, for the SGVM example above, the
crystal is sufficiently short that the agreement is very good though
the discrepancy is somewhat more apparent than for the AGVM example
[see Fig.~\ref{Fig:wigner}(E)]. Finally, for the third example
shown, for which no particular GVM condition is fulfilled, the
disagreement is more apparent. In this case while the analytic and
numerical calculations both predict $\Delta \omega=50.1$THz, the
time durations obtained differ as follows: $\Delta t_{an}=127.2$fs
and $\Delta t_{nu}=205.7$fs [see Fig.~\ref{Fig:wigner}(F)].


We have shown that a factorable two-photon state leads to heralded
single photons which are both pure and FT limited.  We now address
the question of what the relationship is, in general, between the
Schmidt number $K$ (which quantifies the departure from photon pair
factorability) and the time-bandwidth product $\mbox{TB}$ (which
quantifies the departure for the single photons from the FT limit).
Under the Gaussian approximation for the phasematching function it
is possible to carry out an analytical Schmidt decomposition of the
two photon state expressed in Eq.~\ref{E:2photstate} (together with
Eq.~\ref{E:JSA}); such a calculation was carried out in
Ref.~\cite{grice01}. For balanced filtering for both photons
($\sigma_g \rightarrow \infty$), it can be readily shown that the
expression for the Schmidt number thus obtained is exactly identical
to the expression we have derived for the time-bandwidth product
$\mbox{TB}$ [see Eq.~\ref{E:TB}].  We therefore complement the
results previously derived that the single photon purity $p$ is the
reciprocal of the Schmidt number\cite{uren05a} and in turn equal to
the visibility $V$ expected in a so called event-ready
Hong-Ou-Mandel (HOM) interferometer\cite{uren04th} (where two
identical heralded single photons, from separate sources, are made
to interfere in a HOM geometry) with the new result that the
time-bandwidth product is also equal to the Schmidt number. In
summary, we arrive at the following equation linking the degree of
spectral entanglement in the photon pairs ($K$) with important
quantities relating to the resulting heralded single photon:

\begin{equation}
K=\mbox{TB}=1/p=1/V\label{E:KTB}.
\end{equation}

While recent experiments demonstrate nearly ideal single photon
heralding efficiencies, the generation of spectrally pure,
FT-limited single photon wavepackets represents a challenge which
must be addressed to make such sources useful for quantum
information processing applications. In this context, the
relationships expressed by Eq.~\ref{E:KTB} are important for the
design of optimal heralded single photon sources.   Eq.~\ref{E:KTB}
relates an important property of the photon pairs used (the degree
of spectral entanglement, or Schmidt number $K$) with crucial
properties of the resulting heralded single photons.  These
relationships suggest possible experimental procedures to determine
source quality. Specifically, a measurement of the visibility in an
event ready HOM interferometer would yield both the purity and the
time-bandwidth product of the interfering single photon wavepackets.

\section{CONCLUSIONS}

We have studied within the chronocyclic Wigner function (CWF)
formalism the temporal (spectral) properties of heralded single
photon sources based on the process of parametric downconversion,
specifically in a type-II, collinear, frequency-degenerate regime.
We have established, through a Gaussian approximation of the
phasematching function, a clear relationship between the properties
of the photon pairs generated by collinear, degenerate type-II PDC
and those of the resulting heralded single photon.  Despite the
Gaussian approximation used, for an important subset of possible
states, the CWF predicted analytically is essentially identical to
that obtained numerically without recourse to approximations.  We
have shown that spectral entanglement in the photon pairs, as
quantified by the Schmidt number, translates into spectral mixedness
and a departure from the FT-limit in the heralded single photons.
Conversely, we have shown that factorable two photon states lead to
both spectrally pure and Fourier transform limited heralded single
photons.  We have furthermore discussed two distinct group velocity
matching scenarios both of which lead to factorable two photon
states and therefore to pure and Fourier transform limited heralded
single photons, without recourse to filtering.  We have shown that
heralded single photons obtained via asymmetric group velocity
matching lead to the shortest possible single photon time duration,
which corresponds to the pump pulse duration.  The former is
achieved, however, at the cost of a relatively large correlation
time and of distinguishable signal and idler photons. Alternatively,
heralded single photons based on symmetric group velocity matching
lead to a single photon duration greater than the ideal value by a
factor of $\sqrt{2}$ while exhibiting signal-idler
indistinguishability and the lowest possible FT-limited correlation
time.  We believe these results may be important for the further
development of practical single photon sources for quantum
information processing.

\begin{acknowledgements}
We acknowledge useful conversations with I.A. Walmsley, K. Banaszek,
C. Silberhorn and K.A. O'Donnell.  This work was supported by
Conacyt grant 46370-F.
\end{acknowledgements}


\begin{thebibliography}{99}
\bibitem{kok05}See for example review:
P. Kok, W.J. Munro, K. Nemoto, T.C. Ralph, J.P. Dowling, and G.J.
Milburn, quant-ph/0512071.

\bibitem{gisin02} See for example review: N. Gisin, G.G. Ribordy, W.
Tittel and H. Zbinden, Rev. of Mod. Phys. \textbf{74}, 145 (2002)

\bibitem{uren04} A.B. U'Ren, C. Silberhorn, K. Banaszek and I.A. Walmsley, Phys. Rev. Lett. \textbf{93}, 093601 (2004)

\bibitem{uren05b}A.B. U'Ren, J. Ball, Ch. Silberhorn, K. Banaszek and
I. A. Walmsley, Phys. Rev. A \textbf{72} 021802(R) (2005)

\bibitem{pittman05} T.B. Pittman, B.C. Jacobs and J.D. Franson, Opt. Comm. \textbf{246}, 545 (2005)

\bibitem{alibart04} O. Alibart,  D.B. Ostrowsky, P. Baldi and S. Tanzilli, quant-ph/0405075

\bibitem{fasel04} S. Fasel, O. Alibart, S. Tanzilli, P. Baldi, A. Beveratos,
N. Gisin and H. Zbindentext, New J. Phys. \textbf{6}, 163 (2004)

\bibitem{michler00}P. Michler, A. Kiraz, C. Becher, W. V. Schoenfeld, P. M. Petroff, Lidong Zhang, E. Hu, and A. Imamoglu, Science \textbf{290}, 2282
(2000); Z. Yuan, B. E. Kardynal, R. M. Stevenson, A. J. Shields, C.
J. Lobo, K. Cooper, N. S. Beattie, D. A. Ritchie and M. Pepper
Science \textbf{295}, 102 (2002); C. Santori, D. Fattal, J. V.
Caronkovi, G. S. Solomon and Y. Yamamoto, Nature \textbf{419}, 594
(2002)

\bibitem{kuhn02}A. Kuhn, M. Hennrich and G. Rempe, Phys. Rev. Lett. \textbf{89}, 067901 (2002)

\bibitem{kurtsiefer00}C. Kurtsiefer, S. Mayer, P. Zarda and H. Weinfurter,  Phys. Rev. Lett. \textbf{85}, 290 (2000);

\bibitem{rubin00}M.H. Rubin, Phys. Rev. A \textbf{61} 022311 (2000)

\bibitem{cklaw00}C.K. Law, I.A. Walmsley and J. H. Eberly, Phys. Rev.
Lett. \textbf{84}, 5304 (2000)


\bibitem{uren05a} A.B. U'Ren, Ch. Silberhorn, R. Erdmann, K. Banaszek,
W.P. Grice, I.A. Walmsley and M.G. Raymer, Las. Phys. \textbf{15}
146 (2005)



\bibitem{paye92} J. Paye,  Quantum Electronics, IEEE J. Quant. Elect. \textbf{28},
2262 (1992)

\bibitem{walmsley96} I. A. Walmsley and V. Wong, J. Opt. Soc. Am. B \textbf{13}, 2453
(1996)

\bibitem{grice97th} W.P. Grice, Ph.D. thesis, University of Rochester (1997)

\bibitem{friberg85}S. Friberg, C.K. Hong and L. Mandel, Phys. Rev.
Lett. \textbf{54}, 2011 (1985)

\bibitem{note}In this case the CWF is not given by Gaussian
functions, and therefore  $\Delta t \Delta \omega \ge 1$ cannot be
directly applied.

\bibitem{zhang06} L. Zhang, A.B. U'Ren, R. Erdmann, K.A. O'Donnell,
C. Silberhorn, K. Banaszek, I.A. Walsmley, to appear in J. Mod. Opt.

\bibitem{grice01} W. P. Grice, A. B. U'Ren and I. A. Walmsley, Phys. Rev. A
\textbf{64}, 063815 (2001)

\bibitem{factorable} V. Giovannetti, L. Maccone, J. H. Shapiro, and F. N. C. Wong, Phys. Rev. Lett, \textbf{88}, 183602 (2002); A.B. U'Ren, K. Banaszek and I. A. Walmsley Quantum Information and Computation \textbf{3}, 480 (2003); Z.D.
Walton, A. V. Sergienko, B. E. A. Saleh, and M. C. Teich, Phys. Rev.
A \textbf{70}, 052317 (2004); J.P. Torres, F. Maci\`a, S. Carrasco
and L. Torner, Opt. Lett. \textbf{30} 314 (2005); M.G. Raymer, J.
Noh, K. Banaszek and I.A. Walmsley, Phys. Rev. A \textbf{72}, 023825
(2005)

\bibitem{keller97}T.E. Keller and M. H. Rubin, Phys. Rev. A, \textbf{56}, 1534 (1997)

\bibitem{kuzucu05} O. Kuzucu, M. Fiorentino, M. A. Albota, F. N. C. Wong, and F. X. K\"artner, Phys. Rev. Lett. \textbf{94}, 083601 (2005)

\bibitem{uren04th} A.B. U'Ren, Ph.D. thesis, University of Rochester (2004)






\end{thebibliography}

\end{document}